\documentclass{article}
\usepackage[preprint]{spconf,amsmath,graphicx}
\usepackage{hyperref}
\usepackage{booktabs}
\usepackage{wrapfig}
\usepackage{amssymb}
\usepackage{bbm}
\usepackage{url}
\usepackage{float}
\usepackage{pifont}
\usepackage{microtype}
\usepackage{multirow}
\usepackage{color}
\usepackage{cite}
\usepackage{bm}

\copyrightnotice{\copyright\ IEEE 2020}
                \toappear{To appear in {\it Proc.\ ICASSP2020,
                    May 04-08, 2020, Barcelona, Spain}}

\newcommand{\compactand}{\hskip 0.7ex}

\title{Generating diverse and natural text-to-speech samples using a quantized fine-grained VAE and autoregressive prosody prior}

\name{\hspace{-1.3ex}\parbox{\linewidth}{\centering Guangzhi Sun$^{1}$\sthanks{Work performed while interning at Google Brain.}
    \compactand 
    Yu Zhang$^{2}$
    \compactand
    Ron J. Weiss$^{2}$
    \compactand
    Yuan Cao$^{2}$
    \compactand
    Heiga Zen$^{2}$
    \compactand
    Andrew Rosenberg$^{2}$ \protect\\
    Bhuvana Ramabhadran$^{2}$
    \compactand
    Yonghui Wu$^{2}$
    }}
\address{
    $^1$University of Cambridge\hspace{4ex} $^2$Google Inc.\\
    \texttt{\normalsize gs534@eng.cam.ac.uk,\{ngyugh,ronw,yuancao,heigazen,rosenberg,bhuv,yonghui\}@google.com}\\
    }
\begin{document}
\ninept
\maketitle
\begin{abstract}
Recent neural text-to-speech (TTS) models with fine-grained latent features enable precise control of the prosody of synthesized speech.
Such models typically incorporate a fine-grained variational autoencoder (VAE) structure, extracting latent features at each input token (e.g., phonemes).
However, generating samples with the standard VAE prior 
often results in unnatural and discontinuous speech, with dramatic prosodic variation between tokens.
This paper proposes a sequential prior in a discrete latent space which can generate more naturally sounding samples.
This is accomplished by discretizing the latent features using vector quantization (VQ), and separately training an autoregressive (AR) prior model over the result.
%
We evaluate the approach using listening tests, objective metrics of automatic speech recognition (ASR) performance, and measurements of prosody attributes. 
Experimental results show that the proposed model 
significantly improves the naturalness in random sample generation.
Furthermore, initial experiments demonstrate that randomly sampling from the proposed model 
can be used as data augmentation to improve the ASR performance.
\end{abstract}
\section{Introduction}
\label{sec:intro}

The fast-paced development of neural end-to-end TTS synthesis has enabled the generation of speech approaching human levels of naturalness \cite{char2wav,tacotron,deepvoice, taco2}. Such models directly map an input text  to a sequence of acoustic features using an encoder-decoder network architecture \cite{seq2seq}.
In addition to the input text, there are many other sources of variation in speech, including speaker, background noise, channel properties (e.g., reverberation), and prosody.
These attributes can be accounted for in the synthesis model by learning a latent representation as an additional input to the decoder \cite{transfer, styletoken, hsu2018hierarchical, gan4tts}.

Prosody \cite{prosody} collectively refers to stress, intonation and rhythm in speech. Annotations for such factors are rarely available for training.
Many recent end-to-end TTS models aiming to capture these factors extract a latent representation from the target speech, and factorize the observed attributes, such as speaker and text information, out of the prosody latent space \cite{transfer, hsu2018hierarchical, adverserial, battenberg2019effective}.
These approaches extract a single latent variable for an entire utterance, requiring a single global representation to capture the full space of variation across speech signals of arbitrary length.
In contrast, the model proposed in \cite{finegrained} uses a fine-grained structure to encode the 
prosody associated with each phoneme in the input sequence from the aligned target spectrogram.
This system can synthesize speech which closely resembles the prosody of a provided reference speech and control local prosody by varying the values of corresponding latent features. 

There exist applications where generating samples of natural speech corresponding to the same text with different prosody is desirable.
For example, samples with diverse prosodic variations could be useful in data augmentation for ASR where a limited amount of real speech is available.
Using a generative framework, such as VAE \cite{kingma2013vae}, to represent the fine-grained latent variable, naturally enables sampling of different prosody features for each phoneme.
The prior over each latent variable is commonly modeled using a standard Gaussian distribution $\mathcal{N}(\mathbf{0},\mathbf{I})$.  Since the prior is independent at each phoneme, the generated audio often exhibits discontinuous and unnatural artifacts such as long pauses between syllables or sudden increases in energy or fundamental frequency ($F_0$). 

A simple way to ameliorate these unnatural samples is to scale down the standard deviation of the prior distribution during generation, which decreases the likelihood of sampling outlier values.
However, this also suppresses the diversity of the generated audio samples and does not eliminate discontinuities since consecutive samples are still independent.
Attempts to introduce temporal correlation to sequential latent representations \cite{vrnn, vrae} often adopt autoregressive decomposition of the prior and posterior distributions, parameterizing both using neural networks.
More recently, \cite{vqvae} introduced the vector-quantized VAE (VQ-VAE), and  a two-stage training approach for generate high fidelity speech samples, in which the posterior is trained for reconstruction and an autoregressive (AR) prior is trained separately to fit the posteriors extracted from the training set. 

This paper utilizes a two-staged training approach similar to \cite{vqvae}. 
We extend Tacotron~2 \cite{taco2} to incorporate a quantized fine-grained VAE (QFVAE) where the latent representation is quantized into a fixed number of classes.
We find that using a quantized representation improves the naturalness over audio samples generated from the continuous latent space, while still ensuring reasonable diversity across samples.
In the first stage, the TTS model is trained in a teacher-forced setting to maximize the likelihood of the training set.
In the second stage, an autoregressive (AR) \emph{prior network} is trained to fit the VAE posterior distribution over the training data learned in the first stage.
This network learns to model the temporal dynamics across latent features.
Samples characterized by latent features can be drawn from the AR prior by providing an initial state.
We compare two AR prior fitting schemes, one in the continuous space and another in the quantized latent space.

We 
evaluate the proposed model from a few different perspectives. Sample naturalness and completeness are measured using an ASR system trained on real speech data in addition to subjective listening tests evaluating the naturalness of the generated speech. 
Sample diversity is evaluated by the average standard deviation per phoneme in three measurable prosody attributes. Lastly, its benefit as a data augmentation method is demonstrated by the ASR system trained on audio samples generated from TTS systems.


\section{Quantized Fine-grained VAE TTS Model}
\label{sec:fine}
\begin{figure}[t]
\centering
\includegraphics[scale=0.42,trim={0 0.35cm 0 0},clip]{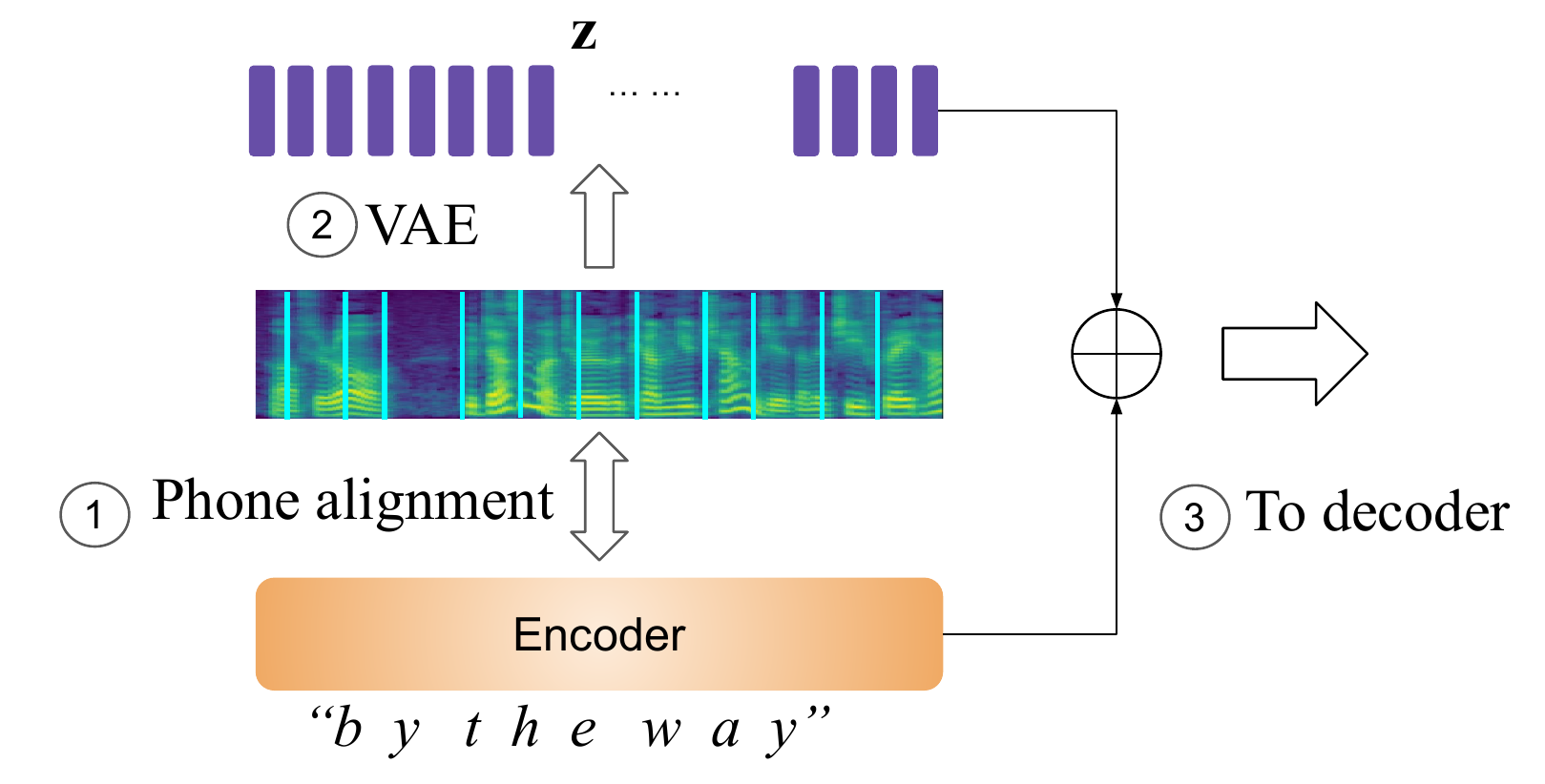}
\caption{Fine-grained VAE encoder structure. Location-sensitive attention is used to align a reference spectrogram with the encoded phoneme sequence. The latent features $\mathbf{z}$ and encoded phonemes are concatenated and passed to the decoder.}
\label{fig:finegrain}
\end{figure}

The fine-grained VAE structure used to model the prosody at phoneme-level, similar to that in \cite{finegrained}, is shown in Fig.~\ref{fig:finegrain}. The VAE component is integrated with the encoder of the Tacotron-2 model \cite{taco2} and the target spectrogram is provided as an extra input to the encoder in order to extract latent  prosody features.

The spectrogram is first aligned with the phoneme sequence using attention \cite{transformer} according to phoneme encodings from the output of the encoder. The aligned spectrogram is then sent to the VAE to extract a sequence of latent representations for the prosody which is also aligned with the phoneme sequence. Finally, the phoneme encodings are concatenated with the latent representations and sent to the decoder. The system is trained by optimizing the fine-grained  evidence lower bound (ELBO) loss:
\begin{multline}
\mathcal{L}(p,q) = \mathbb{E}_{q(\mathbf{z} \mid \mathbf{X})}\left[\log  p(\mathbf{X} \mid \mathbf{Y}, \mathbf{z})\right]  \\
 - \beta \, \textstyle\sum_{n=1}^N D_{\mathrm{KL}}\!\left(q(\mathbf{z}_n \mid \mathbf{X}, \mathbf{Y}_n)\parallel p(\mathbf{z}_n)\right),
\label{eq:elbo}
\vspace{-0.2cm}
\end{multline}
where the first term is the reconstruction loss and the second term is the KL divergence between prior and posterior. The prior is chosen to be $\mathcal{N}(\mathbf{0},\mathbf{I})$. $\mathbf{z}$ represents the sequence of latent features and $\mathbf{z}_n$ corresponds to the latent representation for the $n$-th phoneme. $\mathbf{X}$ is the aligned spectrogram and $\mathbf{Y}$ represents the phoneme encoding. 

\subsection{Vector Quantization}
\label{sec:quant}
\begin{figure}[t]
\centering
\includegraphics[scale=0.45]{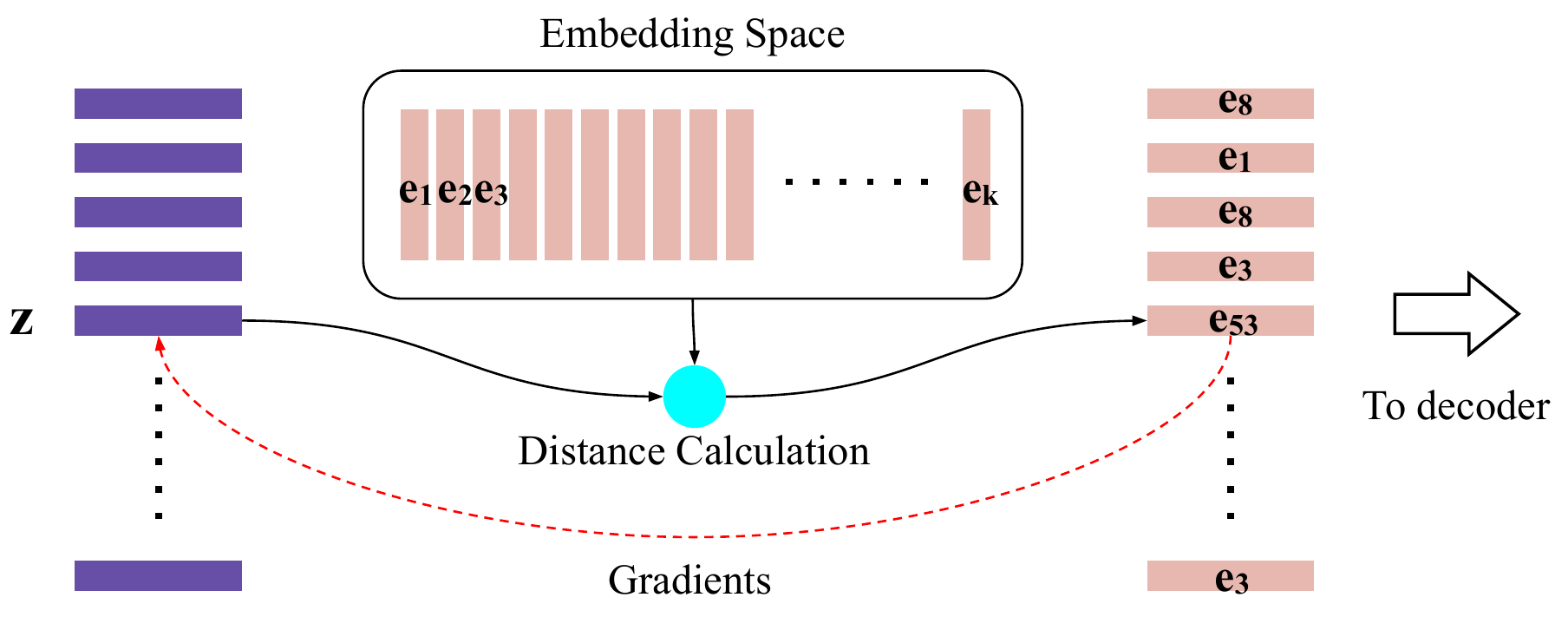}
\caption{Illustration of the vector quantization process. The embedding codebook contains $K$ different embeddings $\mathbf{e}_k$. Continuous latents $\mathbf{z}$ are quantized to their nearest neighbors in the codebook by Euclidean distance. The red line shows the straight-through gradient path.}
\label{fig:vqvae}
\vspace{-2ex}
\end{figure}

Vector quantization in Fig. \ref{fig:vqvae} is performed after the latents are drawn from the posterior distribution by assigning a quantized embedding $\mathbf{e}_k$ to each latent vector $\mathbf{z}_n$ by minimizing the Euclidean distance.
\begin{equation}
k = \text{argmin}_j ||\mathbf{z}_n - \mathbf{e}_j||^2_2
\label{quant}
\end{equation}

Unlike the original VQ-VAE \cite{vqvae} which used a one-hot posterior distribution, we maintain the Gaussian form of the posterior from the standard VAE to make it possible to experiment with continuous or discrete representations within AR prior.
Therefore the phoneme-level ELBO loss is used to train the VQ-VAE as before. 
The gradient from the reconstruction loss term is back-propagated to the latent encoder by directly copying
the gradient from previous layers to quantized embeddings at each step to the latent vectors in the continuous space. Furthermore, to update the quantized embeddings, the following quantization and commitment losses are optimized together with the ELBO:
\begin{equation}
\mathcal{L}_{\mathrm{VQ}} = \textstyle\sum_{n=1}^N\Big{(}||sg[\mathbf{z}_n] - \mathbf{e}_{n}||^2_2 + \gamma ||\mathbf{z}_n - sg[\mathbf{e}_n]||^2_2\Big{)},
\label{quantloss}
\end{equation}
where $sg[\cdot]$ is the stop-gradient operator, which is identity for the forward path and has zero partial derivatives for the backward path. The first term is the \emph{quantization loss} which moves the embeddings towards the latent vectors computed by the encoder, in order to minimize the error introduced in quantization. The second term is the \emph{commitment loss} which encourages the continuous latent vector to remain close to the quantized embedding, preventing the embedding space from expanding too fast. The total loss optimized by the model is the sum of the VAE loss $\mathcal{L}(p, q)$ and the VQ loss $\mathcal{L}_{\mathrm{VQ}}$.

\section{Autoregressive prosody prior}
\label{sec:arprior}
Once the TTS model is trained, the fine-grained VAE encoder shown in Fig.~\ref{fig:finegrain} can be used to compute parameters for the posterior distribution over latents from a reference spectrogram.
Since the posterior is derived from a real speech spectrogram, samples from it will  be natural and coherent across phonemes.
However, without such a reference spectrogram, the model does not expose a method to generate natural samples.
We therefore train an AR prior to model such temporal coherency in the latent feature sequence from the posterior.
This AR prior is trained separately without affecting the training of the posterior. Because the encoder network uses vector quantization after sampling from an underlying posterior distribution in the continuous latent space, we compare two different prior fitting schemes.

The first scheme aims to fit an Gaussian AR prior in the continuous latent space so that the prior and the posterior at each time step come from the same family of distributions. A single layer LSTM is used to model the prior, and is trained using teacher forcing from the latent feature sequence from the posterior. The output at each step of the sequence is a diagonal Gaussian distribution whose mean and standard deviation are functions of the previous latent features. This prior is also conditioned on the phoneme encoding $\mathbf{Y}$:
\begin{equation}
p(\mathbf{z}_n \mid \mathbf{z}_{<n}, \mathbf{Y}) = \mathcal{N}\left(\mathbf{z}_n ; \bm{\mu}(\mathbf{z}_{<n}, \mathbf{Y}), \bm{\sigma}(\mathbf{z}_{<n}, \mathbf{Y})\right),
\label{eq:arcont}
\end{equation}
where $\bm{\mu}(\mathbf{z}_{<n}, \mathbf{Y})$ and $\bm{\sigma}(\mathbf{z}_{<n}, \mathbf{Y})$ are outputs of the prior LSTM.
The LSTM is trained by additionally minimizing the same KL divergence in the continuous latent space as Eq.~\eqref{eq:elbo} except that the original prior $p(\mathbf{z}_n)$ is replaced with $p(\mathbf{z}_n \mid \mathbf{z}_{<n}, \mathbf{Y})$. During generation, the network is only provided with the phoneme encoding and an all-zero initial state. Samples drawn from the continuous AR prior for each phoneme are quantized using the same embedding space. 

Alternatively, the AR prior can be fit directly in the discrete latent space, in which case the prior at each step is a categorical distribution over the quantization embeddings space $\{\mathbf{e}_k\}$.  This is similar to training a neural language model \cite{mikolov2010recurrent}. To enable training the prior with KL divergence in the discrete space, single sample estimation of the posterior distribution is used, hence the training objective becomes the cross-entropy loss:
\vspace{-0.2cm}
\begin{equation}
     D_{\mathrm{KL}}\left(Q(\mathbf{e}) \parallel P(\mathbf{e})\right) = E_{Q(\mathbf{e})}\bigg[\log\frac{Q(\mathbf{e})}{P(\mathbf{e})}\bigg] \approx -\log P(\mathbf{e}_k),
     \label{categorical}
     \vspace{-0.2cm}
\end{equation}
where $Q(\mathbf{e})$ and $P(\mathbf{e})$ are the discrete posterior and prior multinomial distributions, respectively, and $\mathbf{e}_k$ is a single sample drawn from the posterior distribution. Hence the estimated posterior distribution is one-hot at embedding $\mathbf{e}_k$ (i.e.\ $\tilde{Q}(\mathbf{e}_j) = 1$ only when $j=k$, and $0$ otherwise) which yields the cross-entropy loss form. As before, a single layer LSTM is used to model the prior and the output at each step is a categorical distribution after a softmax activation function. The phoneme encoding $\mathbf{Y}$ is also used during training and generation.

\section{Experiments}
\label{sec:exp}

We evaluated multispeaker TTS models on the LibriTTS dataset~\cite{zen2019libritts}, a multispeaker English corpus of approximately 585 hours of read audiobooks sampled at 24kHz. It covers a wide range of speakers, recording conditions and speaking styles. We used a model following \cite{hsu2018hierarchical}, replacing the Gaussian mixture VAE (GMVAE) with a QFVAE.
Output audio is synthesized using a WaveRNN vocoder \cite{kalchbrenner2018efficient}.

To evaluate how much information was lost by quantizing the latent prosody representation, we measured reconstruction performance by encoding a reference signal and using the resulting posterior to reconstruct the same signal.  
In addition, we measured the naturalness of the synthesized speech using subjective listening tests, and speech intelligibility using speech recognition performance with a pretrained ASR model.
We also evaluated the diversity of the prosody in samples from different models.
A good system should be able to generate natural audio samples with a relatively large prosody diversity.

Finally, we demonstrated that samples from the proposed model could be used to augment real speech training data to improve performance of a speech recognition system.
We encourage readers to listen to audio examples on the accompanying web page\footnote{
\url{https://google.github.io/tacotron/publications/prosody_prior}
}.

\subsection{Reconstruction Performance}
Reconstruction, or copy synthesis, performance is measured using $F_0$ frame error (FFE) \cite{ffe} and mel-cepstral distortion (MCD) \cite{mcd} computed from the first 13 MFCCs, which reflect how well a model captures the pitch and timbre of the original speech, respectively.  Lower values are better for both metrics.  Results are shown in Table~\ref{ffe_tab}, illustrating that quantizing the prosody latents in the QFVAE degrades reconstruction compared to the baseline, since some information is discarded as part of the quantization process.
\begin{table}[t]
\centering
\begin{tabular}{lrr}\\\toprule  
\textbf{Model}  & \textbf{FFE} & \textbf{MCD} \\\midrule
Global VAE & 0.52 & 16.0\\
Baseline fine-grained VAE & 0.18 & 8.6\\
\toprule
QFVAE 32 classes & 0.32 & 11.3\\ 
QFVAE 256 classes & 0.26 & 10.2 \\
QFVAE 1024 classes & 0.22 & 9.5\\
\bottomrule
\end{tabular}
\caption{Reconstruction performance for different TTS models. The number of classes refers to the number of latent embeddings in the VQ codebook. The global VAE baseline uses a single latent vector to represent the prosody of an utterance.} 
\label{ffe_tab}
\vspace{-1ex}
\end{table}

Compared to the global VAE model, which used a single 32-dimensional latent vector to represent the prosody across a full utterance, the baseline fine-grained VAE used a 3-dimensional latent for each phoneme and achieves a significantly better reconstruction performance. The QFVAE models have higher FFE and MCD than the baseline, however their performances improve as the number of classes increased.  With 1024 classes 
QFVAE performance approaches that of the baseline. 

\begin{table}[t]
\centering
\addtolength{\tabcolsep}{-2.5pt}
\begin{tabular}{l@{\hspace{0.25em}}lcrrr@{\hspace{0.3em}}r}
\toprule  
&&&&\multicolumn{3}{c}{\textbf{Prosody stddev}} \\
\textbf{Model} & \textbf{Prior} & \textbf{MOS} & \textbf{WER} & 
\textbf{E} & $\mathbf{F_0}$ & \textbf{Dur.}
\\
\midrule
\multicolumn{2}{l}{Real speech} & & 7.2\% & - & - & - \\
\midrule
Baseline & Indep. scale=1.0 & -  & 94.7\% & 0.50 & 58 & 35\\   
Baseline & Indep. scale=0.2 & 2.80 $\pm$ 0.08 & 26.8\% & 0.35 & 38 & 16\\
Baseline & Indep. scale=0.0 & 4.04 $\pm$ 0.06 & 10.6\% & 0.09 & 8 & 8\\
\midrule
Baseline & AR continuous & 2.37 $\pm$ 0.07 & 32.0\% & 0.48 & 32 & 21\\
QFVAE & Indep. &3.43 $\pm$ 0.06 & 11.9\% & 0.38 & 32 & 24 \\
QFVAE & AR discrete & 3.45 $\pm$ 0.07 & 11.5\% & 0.39 & 32 & 16\\
QFVAE & AR continuous & \bf{3.98 $\pm$ 0.06} & 8.4\% & 0.29 & 24 & 12 \\
\bottomrule
\end{tabular}
\caption{Naturalness and diversity metrics computed across samples from different models.  All models use a 3-dimensional fine-grained VAE 
and sample independently from a continuous latent at each step (Indep.) or from an autoregressive (AR) prior over continuous or discrete latents.
The scale factor for the baseline 
scales the standard deviation of the prior when sampling. Prosody metrics include relative energy within each phonene (E), fundamental frequency in Hertz ($F_0$), and duration in ms (Dur.).}
\label{tab:sample}
\vspace{-2ex}
\end{table} 


\subsection{Sample Naturalness and Diversity}

We conducted subjective listening tests over 10 speakers each synthesized 100 utterances with native English speakers asked to rate the naturalness of speech samples on a 5-point scale in increments of 0.5. Results are reported in terms of mean opinion score (MOS).

Complementary to naturalness, we also report the word error rate (WER) from an ASR model trained on real speech from the LibriTTS training set and evaluated on speech synthesized from transcripts in the LibriTTS test set\footnote{WER on LibriTTS is different from WER on LibriSpeech.}. We used the sequence-to-sequence ASR model from \cite{irie2019choice}. 
This metric verifies that the synthesized speech contains the full content of the input text.
However, even if the full text is synthesized, we expect the WER to increase for speech samples with prosody that is very inconsistent with that of real speech.

Finally, we evaluated the diversity of samples from the proposed prior by measuring the standard deviation of three prosody attributes computed for each phoneme: relative energy, fundamental frequency ($F_0$), and duration.
The duration of a phoneme is represented by the number of frames where the decoder attention assigns maximum values to that phoneme, $F_0$ is computed by the YIN pitch tracker \cite{yin}, and the relative energy is the ratio of the average signal magnitude within a phoneme with the average magnitude of the entire utterance.
The standard deviations for each of these metrics is computed within each phoneme, and the average standard deviation across all the phonemes is reported.  
To estimate these statistics, we synthesized 100 random samples for each of 3 randomly selected utterances from the test set, each using 3 different speaker IDs. 

Results are shown in Table~\ref{tab:sample}. 
Comparing independent sampling strategies from the baseline model, decreasing the scale reduced the diversity in each attribute, while increasing naturalness MOS and improving WER. This indicates a trade-off between the diversity and naturalness that results from naive sampling from the baseline system.
A scaling factor of $0.2$ is the largest under which the system still generated somewhat natural audio samples.
Samples using a scale of $1.0$ were too poor (as reflected in the very high WER) that we did not conduct listening tests with them.
%
Fitting an AR prior over the baseline's continuous latent space results in worse naturalness than sampling independently for each phoneme with moderate scale of $0.2$, although the diversity becomes higher.

%
QFVAE samples always result in reasonable MOS and WER regardless of the type of prior.
This indicates the benefit of the regularization imposed by quantizing the latent space during training, even if the discrete representation is not used directly by the prior.
The most natural results with highest MOS and lowest WER resulted from using an AR prior fit in the continuous space.  
Samples generated using this prior have MOS close to the baseline with neutral prosody ($\text{scale}=0.0$), but with lower WER.
%
However independent samples had better diversity metrics, once again reflecting a similar, but less pronounced, diversity-quality trade-off to the baseline model.
Finally, the AR prior in the discrete space gives a similar diversity and naturalness to the independent prior.
We conjecture that using a single sample to estimate the discrete KL divergence brings uncertainty to the model, resulting in a prior with large variance at each step.

\subsection{Data Augmentation}

One potential application of the proposed TTS model 
is to sample synthetic speech to help training ASR models, in a data augmentation procedure.
We trained ASR models on synthesized audio from different TTS models and evaluated how well the resulting recognizers generalized when evaluated on real speech.

For each TTS model, we synthesized speech for the full set of training \emph{transcripts} in LibriTTS with randomized speaker IDs.
The synthesized speech was downsampled to a rate of 16 kHz and then used to train the ASR model.
We used an end-to-end encoder-decoder ASR model with additive attention \cite{chiu2018state}.
The model, training strategy, and associated hyperparameters followed LAS-4-1024 in \cite{park2019specaugment}. 

As shown in Table~\ref{tab:sample_tts4asr}, the WER of the ASR model when trained on real LibriTTS speech is 7.2\%.
Similar to \cite{li2018training}, synthesizing speech using a baseline multispeaker Tacotron~2 model results in a significantly degraded WER of 20.0\%, indicating that samples from this model did not capture the full space of variation of real speech.

In a copy synthesis setting where the ground truth reference speech is provided, the global VAE \cite{hsu2018hierarchical} improves on the Tacotron baseline, and the fine-grained VAE (Baseline) improves even further. 
These results indicate the importance of explicitly modeling prosody variation (VAE models) as well as speaker variation (Tacotron~2).


Randomly sampling independently at each phoneme using the baseline fine-grained VAE performs significantly worse than copy synthesis. The lowest WER is obtained when the scale is set to an intermediate value, reflecting a reasonable trade-off between naturalness and diversity.  
As in Table~\ref{tab:sample}, the best QFVAE performance results from sampling independently at each phoneme.


To explore the improvement from diversity, the best performing QFVAE was used to generate 10 copies with varying prosody of the training data to train the ASR system. This gives 12.5\% WER which is close to the copy synthesize result. Negligible improvement is found when adding more copies for Tacotron 2 model since they do not contain much variation in prosody.
Finally, we found that training on synthetic speech (oversampled ten times) with real speech in a data augmentation configuration resulted in a 16\% relative WER reduction compared to the model trained on real speech alone.


\begin{table}[t]
\centering
\begin{tabular}{llr}\\
\toprule  
\textbf{Model} & \textbf{Prior} & \textbf{WER} \\\midrule
\multicolumn{2}{l}{Real speech} &  7.2\%\\
\multicolumn{2}{l}{Multispeaker Tacotron 2} &  20.0\% \\
\multicolumn{2}{l}{GMVAE-Tacotron \cite{hsu2018hierarchical} copy synthesis} &  16.4\% \\
\multicolumn{2}{l}{Baseline copy synthesis} &  11.8\%\\
\midrule
Baseline & Indep. scale=1.0 & 25.6\% \\   
Baseline & Indep. scale=0.2 & 19.9\% \\
Baseline & Indep. scale=0.0 & 31.3\% \\
Baseline & AR continuous & 32.0\% \\
QFVAE & Indep. & \textbf{16.9}\%  \\
QFVAE & AR discrete & 17.6\%  \\
QFVAE & AR continuous & 22.3\% \\
\midrule
\multicolumn{3}{l}{\hspace{-1.25ex}10 samples per transcript} \\
\multicolumn{2}{l}{Tacotron 2} & 19.8\% \\
Baseline & Indep. scale=0.2 & 13.8\% \\
QFVAE & Indep. & \textbf{12.5}\% \\
Real speech + QFVAE  & Indep. & \textbf{6.0}\%\\
\bottomrule
\end{tabular}
\caption{WER on real speech from the LibriTTS test set for ASR models trained only on \emph{synthesized} speech sampled from TTS models.}
\label{tab:sample_tts4asr}
\vskip-3ex
\end{table} 

We repeated the evaluation of the data augmentation experiment on the LibriSpeech corpus \cite{panayotov2015librispeech} in Table~\ref{tab:librispeech}. 
The real speech used in this comparison was the union of LibriTTS and LibriSpeech training sets, the TTS model was trained on LibriTTS, which contained material that was not in LibriSpeech.
Using the QFVAE for data augmentation resulted in relative WER reductions of 14\% and 8\% on the LibriSpeech test-clean and test-other sets, respectively.

\begin{table}[t]
\addtolength{\tabcolsep}{-2pt}
\centering
\begin{tabular}{lcc}\\\toprule  
\textbf{Training data}  & \textbf{test} & \textbf{test-other}\\\midrule
Real speech & 4.4\% & 12.4\%\\
Real speech + QFVAE (Indep.) 10 samples & \textbf{3.8}\% & \textbf{11.4}\%\\
\bottomrule
\end{tabular}
\caption{WER on real speech LibriSpeech test sets for ASR models trained on the combination of real and synthesized speech.}
\label{tab:librispeech}
\end{table} 


\section{Conclusions}
\label{sec:conc}

This paper proposed a quantized fine-grained VAE TTS model, and compared different prosody priors to synthesize natural and diverse audio samples. A set of evaluations for naturalness and diversity was provided. Results showed that the quantization improved the sample naturalness while retaining a similar diversity. Sampling from an AR prior further improved the naturalness. When generated samples were used to train ASR systems, we demonstrated a potential application that used prosody variations for data augmentation. 

\section{Acknowledgements}
The authors thank Daisy Stanton, Eric Battenberg, and the Google Brain and Perception teams for their helpful feedback and discussions.
\newpage


\bibliographystyle{IEEEbib}
\bibliography{qfvae}

\begin{thebibliography}{10}

\bibitem{char2wav}
J.~Sotelo, S.~Mehri, K.~Kumar, J.~Santos, K.~Kastner, A.~Courville, and
  Y.~Bengio.,
\newblock ``{Char2Wav}: End-to-end speech synthesis.,''
\newblock in {\em Proc. International Conference on Learning Representations
  (ICLR)}, 2017.

\bibitem{tacotron}
Y.~Wang, R.~J. Skerry-Ryan, D.~Stanton, Y.~Wu, R.~J. Weiss, N.~Jaitly, Z.~Yang,
  Y.~Xiao, Z.~Chen, S.~Bengio, Q.~Le, Y.~Agiomyrgiannakis, R.~Clark, and R.~A.
  Saurous.,
\newblock ``Tacotron: Towards end-to-end speech synthesis.,''
\newblock in {\em Proc. Interspeech}, 2017, pp. 4006--4010.

\bibitem{deepvoice}
W.~Ping, K.~Peng, A.~Gibiansky, S.~O. Arik, A.~Kannan, S.~Narang, J.~Raiman,
  and J.~Miller.,
\newblock ``Deep voice 3: 2000-speaker neural text-to-speech.,''
\newblock in {\em Proc. International Conference on Learning Representations
  (ICLR)}, 2018.

\bibitem{taco2}
J.~Shen, R.~Pang, R.~J. Weiss, M.~Schuster, N.~Jaitly, Z.~Yang, Z.~Chen,
  Y.~Zhang, Y.~Wang, and R.~J. Skerry-Ryan,
\newblock ``Natural {TTS} synthesis by conditioning {WaveNet} on mel
  spectrogram predictions.,''
\newblock in {\em Proc. ICASSP}, 2018, pp. 4779--4783.

\bibitem{seq2seq}
I.~Sutskever, O.~Vinyals, and Q.~V. Le,
\newblock ``Sequence to sequence learning with neural networks,''
\newblock in {\em Advances in Neural Information Processing Systems}, 2014.

\bibitem{transfer}
R.~J. Skerry-Ryan, E.~Battenberg, Y.~Xiao, Y.~Wang, D.~Stanton, J.~Shor,
  R.~Weiss, R.~Clark, and R.~A. Saurous,
\newblock ``Towards end-to-end prosody transfer for expressive speech synthesis
  with {Tacotron},''
\newblock in {\em Proc. International Conference on Machine Learning (ICML)},
  2018.

\bibitem{styletoken}
Y.~Wang, D.~Stanton, Y.~Zhang, R.~J.~S. Ryan, E.~Battenberg, J.~Shor, Y.~Xiao,
  Y.~Jia, F.~Ren, and R.~A. Saurous,
\newblock ``Style tokens: Unsupervised style modeling, control and transfer in
  end-to-end speech synthesis,''
\newblock in {\em Proc. International Conference on Machine Learning (ICML)},
  2018, pp. 5167--5176.

\bibitem{hsu2018hierarchical}
W.-N. Hsu, Y.~Zhang, R.~J. Weiss, H.~Zen, Y.~Wu, Y.~Wang, Y.~Cao, Y.~Jia,
  Z.~Chen, J.~Shen, P.~Nguyen, and R.~Pang,
\newblock ``Hierarchical generative modeling for controllable speech
  synthesis,''
\newblock in {\em Proc. International Conference on Learning Representations
  (ICLR)}, 2019.

\bibitem{gan4tts}
S.~Ma, D.~Mcduff, and Y.~Song.,
\newblock ``A generative adversarial network for style modeling in a
  text-to-speech system,''
\newblock in {\em Proc. International Conference on Learning Representations
  (ICLR)}, 2019.

\bibitem{prosody}
M.~Wagner and D.~G. Watson,
\newblock ``Experimental and theoretical advances in prosody: A review,''
\newblock in {\em Language and Cognitive Processes}, 2010, pp. 905--945.

\bibitem{adverserial}
W.-N. Hsu, Y.~Zhang, R.~J. Weiss, Y.-A. Chung, Y.~Wang, Y.~Wu, and J.~Glass.,
\newblock ``Disentangling correlated speaker and noise for speech synthesis via
  data augmentation and adversarial factorization,''
\newblock in {\em Proc. ICASSP}, 2019.

\bibitem{battenberg2019effective}
E.~Battenberg, S.~Mariooryad, D.~Stanton, R.~J. Skerry-Ryan, M.~Shannon,
  D.~Kao, and T.~Bagby,
\newblock ``Effective use of variational embedding capacity in expressive
  end-to-end speech synthesis,''
\newblock {\em arXiv: 1906.03402}, 2019.

\bibitem{finegrained}
Y.~Lee and T.~Kim,
\newblock ``Robust and fine-grained prosody control of end-to-end speech
  synthesis,''
\newblock in {\em Proc. ICASSP}, 2019, pp. 5911--5915.

\bibitem{kingma2013vae}
D.~P. Kingma and M.~Welling,
\newblock ``Auto-encoding variational bayes,''
\newblock in {\em Proc. International Conference on Learning Representations
  (ICLR)}, 2014.

\bibitem{vrnn}
J.~Chung, K.~Kastner, L.~Dinh, K.~Goel, A.~Courville, and Y.~Bengio,
\newblock ``A recurrent latent variable model for sequential data,''
\newblock in {\em Advances in Neural Information Processing Systems}, 2016.

\bibitem{vrae}
O.~Fabius and J.~R. van Amersfoort,
\newblock ``Variational recurrent auto-encoders,''
\newblock in {\em Proc. International Conference on Learning Representations
  (ICLR)}, 2015.

\bibitem{vqvae}
A.~van~den Oord, O.~Vinyals, and K.~Kavukcuoglu,
\newblock ``Neural discrete representation learning,''
\newblock in {\em Advances in Neural Information Processing Systems}, 2017.

\bibitem{transformer}
A.~Vaswani, N.~Shazeer, N.~Parmar, J.~Uszkoreit, L.~Jones, A.~N. Gomez,
  L.~Kaiser, and I.~Polosukhin,
\newblock ``Attention is all you need,''
\newblock in {\em Advances in Neural Information Processing Systems}, 2017.

\bibitem{mikolov2010recurrent}
T.~Mikolov, M.~Karafi{\'a}t, L.~Burget, J.~{\v{C}}ernock{\`y}, and
  S.~Khudanpur,
\newblock ``Recurrent neural network based language model,''
\newblock in {\em Proc. Interspeech}, 2010.

\bibitem{zen2019libritts}
H.~Zen, V.~Dang, R.~Clark, Y.~Zhang, R.~J. Weiss, Y.~Jia, Z.~Chen, and Y.~Wu,
\newblock ``{LibriTTS}: A corpus derived from {LibriSpeech} for
  text-to-speech,''
\newblock in {\em Proc. Interspeech}, 2019.

\bibitem{kalchbrenner2018efficient}
N.~Kalchbrenner, E.~Elsen, K.~Simonyan, S.~Noury, N.~Casagrande, E.~Lockhart,
  F.~Stimberg, A.~Oord, S.~Dieleman, and K.~Kavukcuoglu,
\newblock ``Efficient neural audio synthesis,''
\newblock in {\em Proc. International Conference on Machine Learning (ICML)},
  2018, pp. 2415--2424.

\bibitem{ffe}
W.~Chu and A.~{Alwan},
\newblock ``Reducing {F0} frame error of {F0} tracking algorithms under noisy
  conditions with an unvoiced/voiced classification frontend,''
\newblock in {\em Proc. ICASSP}, 2009.

\bibitem{mcd}
R.~{Kubichek},
\newblock ``Mel-cepstral distance measure for objective speech quality
  assessment,''
\newblock in {\em Proc. IEEE Pacific Rim Conference on Communications Computers
  and Signal Processing}, 1993, vol.~1, pp. 125--128.

\bibitem{irie2019choice}
K.~Irie, R.~Prabhavalkar, A.~Kannan, A.~Bruguier, D.~Rybach, and P.~Nguyen,
\newblock ``On the choice of modeling unit for sequence-to-sequence speech
  recognition,''
\newblock in {\em Proc. Interspeech}, 2019, pp. 3800--3804.

\bibitem{yin}
A.~de~Cheveign\'{e} and H.~Kawahara,
\newblock ``{YIN}, a fundamental frequency estimator for speech and music,''
\newblock {\em The Journal of the Acoustical Society of America}, vol. 111, no.
  4, pp. 1917--1930, 2002.

\bibitem{chiu2018state}
C.-C. Chiu, T.~N. Sainath, Y.~Wu, R.~Prabhavalkar, P.~Nguyen, Z.~Chen,
  A.~Kannan, R.~J. Weiss, K.~Rao, E.~Gonina, et~al.,
\newblock ``State-of-the-art speech recognition with sequence-to-sequence
  models,''
\newblock in {\em Proc. ICASSP}, 2018, pp. 4774--4778.

\bibitem{park2019specaugment}
D.~S. Park, W.~Chan, Y.~Zhang, C.-C. Chiu, B.~Zoph, E.~D. Cubuk, and Q.~V. Le,
\newblock ``{SpecAugment}: A simple data augmentation method for automatic
  speech recognition,''
\newblock in {\em Proc. Interspeech}, 2019.

\bibitem{li2018training}
J.~Li, R.~Gadde, B.~Ginsburg, and V.~Lavrukhin,
\newblock ``Training neural speech recognition systems with synthetic speech
  augmentation,''
\newblock {\em arXiv preprint arXiv:1811.00707}, 2018.

\bibitem{panayotov2015librispeech}
V.~Panayotov, G.~Chen, D.~Povey, and S.~Khudanpur,
\newblock ``{LibriSpeech}: An {ASR} corpus based on public domain audio
  books,''
\newblock in {\em Proc. ICASSP}, 2015, pp. 5206--5210.

\end{thebibliography}

\end{document}